\documentclass[10pt,conference]{IEEEtran}

\usepackage[dvips]{graphicx}

\newcommand{\bysame}{%
    \leavevmode\hbox to 3em{\hrulefill}\,}

\begin{document}

\title{Error-Trellis State Complexity of LDPC Convolutional Codes Based on Circulant Matrices}

\author{
\authorblockN{Masato Tajima}
\authorblockA{Dept of Intellect. Inform. Systems Eng.  \\
University of Toyama \\
Toyama 930-8555, Japan \\
tajima@eng.u-toyama.ac.jp}
\and
\authorblockN{Koji Okino}
\authorblockA{Information Technology Center \\
University of Toyama \\
Toyama 930-8555, Japan \\
okino@itc.u-toyama.ac.jp}
\and
\authorblockN{Takashi Miyagoshi}
\authorblockA{Dept of Intellect. Inform. Systems Eng.  \\
University of Toyama \\
Toyama 930-8555, Japan \\
miyagosi@eng.u-toyama.ac.jp}
}
%

\maketitle

\begin{abstract}
Let $H(D)$ be the parity-check matrix of an LDPC convolutional code corresponding to the parity-check matrix $H$ of a QC code obtained using the method of Tanner et al. We see that the entries in $H(D)$ are all monomials and several rows (columns) have monomial factors. Let us cyclically shift the rows of $H$. Then the parity-check matrix $H'(D)$ corresponding to the modified matrix $H'$ defines another convolutional code. However, its free distance is lower-bounded by the minimum distance of the original QC code. Also, each row (column) of $H'(D)$ has a factor different from the one in $H(D)$. We show that the state-space complexity of the error-trellis associated with $H'(D)$ can be significantly reduced by controlling the row shifts applied to $H$ with the error-correction capability being preserved.
\end{abstract}

\section{Introduction}
In this paper, we assume that the underlying field is $F=\mbox{GF}(2)$. Let $G(D)$ be a polynomial generator matrix for an $(n_0, k_0, \nu)$ convolutional code $C$ with memory $\nu$. Denote by $H(D)$ a corresponding parity-check matrix. Both $G(D)$ and $H(D)$ are assumed to be canonical [4], [5]. In this case, the code-trellis module associated with $G(D)$ and the error-trellis modules [6] associated with the syndrome former $H^T(D)$ ($T$ means transpose) have $2^{\nu}$ states, where the obvious realization of $G(D)$ and the adjoint-obvious realization [3] of $H^T(D)$ are assumed, respectively. Ariel and Snyders [1] presented a construction of an error-trellis based on the scalar check matrix derived from $H(D)$. They showed that when some ($j$th) ``column'' of $H(D)$ has a factor $D^l$ (i.e., the $j$th column is not ``delay free''), there is a possibility that state-space reduction can be realized. Being motivated by their work, we also examined the same case. We took notice of a syndrome generation process. The time-$k$ error $\mbox{\boldmath $e$}_k$ and the time-$k$ syndrome $\mbox{\boldmath $\zeta$}_k$ are connected with the relation $\mbox{\boldmath $\zeta$}_k=\mbox{\boldmath $e$}_kH^T(D)$. From this relation, we noticed [8] that the transformation $e_k^{(j)}\rightarrow D^le_k^{(j)}=e_{k-l}^{(j)}$ is equivalent to dividing the $j$th column of $H(D)$ by $D^l$. That is, reduction can be accomplished by shifting the ``subsequence'' $\{e_k^{(j)}\}$ of the original error-path.
\par
On the other hand, consider the parity-check matrix
\begin{equation}
H_1(D)\stackrel{\triangle}{=}\left(
\begin{array}{ccc}
D^2& D^2 & 1 \\
1& 1+D+D^2 & 0 
\end{array}
\right) .
\end{equation}
Since $H_1(D)$ is canonical and all the columns are delay free, any further reduction seems to be impossible. In fact, it follows from {\it Theorem 1} of [1] that the dimension $d_1$ of the state space of the error-trellis based on $H_1^T(D)$ is $4$. However, a corresponding generator matrix is given by $G_1(D)\stackrel{\triangle}{=}(1+D+D^2, 1, D^3+D^4)$. Note that the third column of $G_1(D)$ has a factor $D^2$. ({\it Remark:} It suffices to divide the third column by $D^2$ in order to obtain a reduced code-trellis.) This fact implies that a reduced error-trellis can be constructed [1], [8] (i.e., state-space reduction can be realized). Then consider the reciprocal dual encoder [4]
\begin{equation}
\tilde H_1(D)\stackrel{\triangle}{=}\left(
\begin{array}{ccc}
1& 1 & D^2 \\
D^2& 1+D+D^2 & 0 
\end{array}
\right) .
\end{equation}
Note that the third column of $\tilde H_1(D)$ has a factor $D^2$. Accordingly, dividing the third column of $\tilde H_1(D)$ by $D^2$, we can construct an error-trellis with $4$ states (i.e., $\tilde d_1=2$) [1], [8]. Here, notice that each error-path in the error-trellis based on $H_1^T(D)$ can be represented in time-reversed order using the error-trellis based on $\tilde H_1^T(D)$. Hence, a factor $D^2$ in the column of $\tilde H_1(D)$ corresponds to backward-shifting by two time units (i.e., $D^{-2}$) in terms of the original $H_1(D)$. Actually, by ``multiplying'' the third column of $H_1(D)$ by $D^2$, we have
\begin{equation}
H_1'(D)\stackrel{\triangle}{=}\left(
\begin{array}{ccc}
D^2& D^2 & D^2 \\
1& 1+D+D^2 & 0 
\end{array}
\right) .
\end{equation}
Note that this matrix can be reduced to an equivalent canonical parity-check matrix
\begin{equation}
H_1''(D)\stackrel{\triangle}{=}\left(
\begin{array}{ccc}
1& 1 & 1 \\
1& 1+D+D^2 & 0 
\end{array}
\right)
\end{equation}
by dividing the first ``row'' by $D^2$. Hence, the dimension $d_1$ can be reduced to $2$. ({\it Remark:} This fact cannot be derived from the results of [1].) It follows from the above argument that there is a possibility that a reduced error-trellis can be constructed not only using forward-shifted error subsequences but also using backward-shifted error subsequences.
\par
Now, we remark that a parity-check matrix $H(D)$ with the form described above appears in [10]. Tanner et al. [10] presented a class of algebraically constructed quasi-cyclic (QC) LDPC codes and their convolutional counterparts. It is stated that the convolutional codes obtained in the paper typically have large constraint lengths and therefore the use of trellis-based decoding is not feasible. However, the parity-check matrices of LDPC convolutional codes proposed by Tanner et al. have monomial entries. Accordingly, the above-mentioned state-space reduction method can be directly applied to those parity-check matrices. Then we intended to evaluate the state-space complexity of the error-trellis of an LDPC convolutional code which appears in [10]. We show that the {\it overall constraint length} (abbreviated as ``OCL'' in this paper) of the parity-check matrix which specifies an LDPC convolutional code can be significantly reduced with the error-correction capability of the convolutional code being preserved.
\section{Preliminaries}
\subsection{Error-Trellis Construction Using Shifted Error/Syndrome Subsequences}
Let $H(D)$ be a parity-check matrix for an $(n_0, k_0)$ convolutional code $C$. In this paper, we consider the error-trellis based on the syndrome former $H^T(D)$. In this case, the adjoint-obvious realization of $H^T(D)$ is assumed unless otherwise specified. Denote by $\mbox{\boldmath $e$}_k=(e_k^{(1)}, \cdots, e_k^{(j)}, \cdots, e_k^{(n_0)})$ and $\mbox{\boldmath $\zeta$}_k=(\zeta_k^{(1)}, \cdots, \zeta_k^{(i)}, \cdots, \zeta_k^{(r)})$ the time-$k$ error and the time-$k$ syndrome, respectively, where $r=n_0-k_0$. Then we have the relation:
\begin{equation}
\mbox{\boldmath $\zeta$}_k=\mbox{\boldmath $e$}_kH^T(D) .
\end{equation}
Assume that the $i$th row of $H(D)$ has the form
\begin{equation}
\left(
\begin{array}{cccc}
D^{l_i}h_{i1}'(D) &D^{l_i}h_{i2}'(D) &\ldots &D^{l_i}h_{in_0}'(D)
\end{array}
\right) ,
\end{equation}
where $l_i\geq 1$. Let $H'(D)$ be the modified version of $H(D)$ with the $i$th row being replaced by
\begin{equation}
\left(
\begin{array}{cccc}
h_{i1}'(D) &h_{i2}'(D) &\ldots &h_{in_0}'(D)
\end{array}
\right) .
\end{equation}
Defining $\zeta_k^{'(i)}$ as $\zeta_k^{(i)}\stackrel{\triangle}{=}D^{l_i}\zeta_k^{'(i)}=\zeta_{k-l_i}^{'(i)}$, we set $\mbox{\boldmath $\zeta$}_k'\stackrel{\triangle}{=}(\zeta_k^{(1)}, \cdots, \zeta_k^{'(i)}, \cdots, \zeta_k^{(r)})$. Then we have
\begin{equation}
\mbox{\boldmath $\zeta$}_k'=\mbox{\boldmath $e$}_kH^{'T}(D) .
\end{equation}
Similarly, assume that the $j$th column of $H(D)$ has the form
\begin{equation}
\left(
\begin{array}{cccc}
D^{l_j}h_{1j}'(D) &D^{l_j}h_{2j}'(D) &\ldots &D^{l_j}h_{rj}'(D)
\end{array}
\right)^T ,
\end{equation}
where $l_j\geq 1$. ({\it Remark:} $H(D)$ is not basic [2] and then not canonical.) Let $H'(D)$ be the modified version of $H(D)$ with the $j$th column being replaced by
\begin{equation}
\left(
\begin{array}{cccc}
h_{1j}'(D) &h_{2j}'(D) &\ldots &h_{rj}'(D)
\end{array}
\right)^T .
\end{equation}
Also, let $\mbox{\boldmath $e$}_k'\stackrel{\triangle}{=}(e_k^{(1)}, \cdots, e_k^{'(j)}, \cdots, e_k^{(n_0)})$, where $e_k^{'(j)}\stackrel{\triangle}{=}D^{l_j}e_k^{(j)}=e_{k-l_j}^{(j)}$. Then we have
\begin{equation}
\mbox{\boldmath $\zeta$}_k=\mbox{\boldmath $e$}_k'H^{'T}(D) .
\end{equation}
Noting these relations [8], [9], in the case where the $i$th row of $H(D)$ has a factor $D^{l_i}$, by shifting the $i$th syndrome subsequence by $l_i$ time units, whereas in the case where the $j$th column of $H(D)$ has a factor $D^{l_j}$, by shifting the $j$th error subsequence by $l_j$ time units, we can construct an error trellis with reduced number of states. In the following, we call factoring out $D^{l}$ from a row of $H(D)$ and from a column of $H(D)$ ``row operation'' and ``column operation'', respectively.

\subsection{Error-Trellis Construction Based on a Reciprocal Dual Encoder}

\begin{figure}[tb]
\begin{center}
\includegraphics[width=7.5cm,clip]{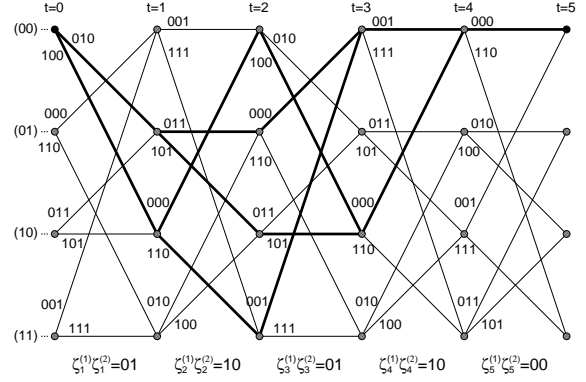}
\end{center}
\caption{Example error-trellis based on $H_2^T(D)$.}
\label{Fig.1}
\end{figure}

Consider the $(3, 1, 2)$ convolutional code {\it $C_2$} with canonical parity-check matrix given by
\begin{equation}
H_2(D)\stackrel{\triangle}{=}\left(
\begin{array}{ccc}
D& 0 & 1 \\
1& 1+D & 0 
\end{array}
\right) .
\end{equation}
In this subsection, we discuss using this specific example. However, the argument is entirely general. Since the columns of $H_2(D)$ are delay free, the dimension $d_2$ of the state space of the error-trellis based on the syndrome former $H_2^T(D)$ is given by $2$ (see {\it Theorem 1} of [1]). Fig.1 shows an error-trellis constructed based on $H_2^T(D)$ using the conventional method [6]. It is assumed that a transmitted code-path is terminated in the all-zero state at $t=4$ and the corresponding received data is given by $\mbox{\boldmath $z$}=\mbox{\boldmath $z$}_1~\mbox{\boldmath $z$}_2~\mbox{\boldmath $z$}_3~\mbox{\boldmath $z$}_4~\mbox{\boldmath $z$}_5=010~011~000~001~000$, where $\mbox{\boldmath $z$}_5=000$ is the imaginary received data. Let {\boldmath $z$} be the input of the syndrome former $H_2^T(D)$, then we have the syndrome sequence $\mbox{\boldmath $\zeta$}=\mbox{\boldmath $\zeta$}_1~\mbox{\boldmath $\zeta$}_2~\mbox{\boldmath $\zeta$}_3~\mbox{\boldmath $\zeta$}_4~\mbox{\boldmath $\zeta$}_5=01~10~01~10~00$. The overall error-trellis is constructed by concatenating five error-trellis modules corresponding to $\mbox{\boldmath $\zeta$}_k$. Note that the error-trellis in Fig.1 is terminated in state $(00)$ at $t=5$, which corresponds to the final syndrome-former state $\mbox{\boldmath $\sigma$}_5=(00)$. From Fig.1 (note that $\mbox{\boldmath $e$}_5=000$), we have four admissible error-paths:
\begin{eqnarray}
\mbox{\boldmath $e$}_{p_1} &=& 010~011~000~001~000 \\
\mbox{\boldmath $e$}_{p_2} &=& 010~101~101~000~000 \\
\mbox{\boldmath $e$}_{p_3} &=& 100~000~100~000~000 \\
\mbox{\boldmath $e$}_{p_4} &=& 100~110~001~001~000 .
\end{eqnarray}


\begin{figure}[tb]
\begin{center}
\includegraphics[width=7.5cm,clip]{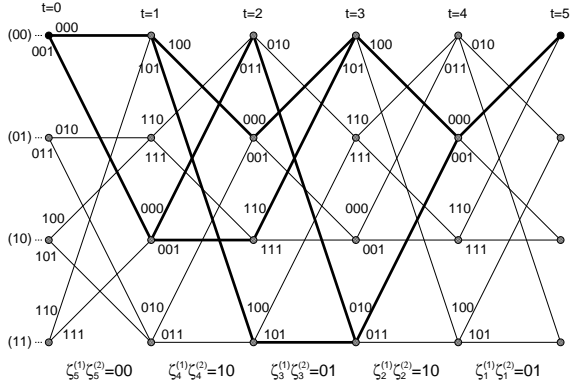}
\end{center}
\caption{Error-trellis based on $\tilde H_2^T(D)$.}
\label{Fig.2}
\end{figure}

Next, consider the reciprocal dual encoder
\begin{equation}
\tilde H_2(D)\stackrel{\triangle}{=}\left(
\begin{array}{ccc}
1& 0 & D \\
D& 1+D & 0 
\end{array}
\right) .
\end{equation}
Let $\mbox{\boldmath $\tilde z$}=\mbox{\boldmath $z$}_4~\mbox{\boldmath $z$}_3~\mbox{\boldmath $z$}_2~\mbox{\boldmath $z$}_1~\mbox{\boldmath $z$}_0=001~000~011~010~000$ be the time-reversed received data of $\{\mbox{\boldmath $z$}_k\}_{k=1}^4$ augmented with the imaginary data $\mbox{\boldmath $z$}_0=000$. If $\mbox{\boldmath $\tilde z$}$ is inputted to the syndrome former $\tilde H_2^T(D)$, then the time-reversed syndrome sequence $\mbox{\boldmath $\tilde \zeta$}=\mbox{\boldmath $\zeta$}_5~\mbox{\boldmath $\zeta$}_4~\mbox{\boldmath $\zeta$}_3~\mbox{\boldmath $\zeta$}_2~\mbox{\boldmath $\zeta$}_1=00~10~01~10~01$ is obtained. The corresponding error-trellis is shown in Fig.2, where the trellis is terminated in state $(00)$, which corresponds to the final syndrome-former state $\mbox{\boldmath $\tilde \sigma$}_5=(00)$. From Fig.2, we have four admissible error-paths:
\begin{eqnarray}
\mbox{\boldmath $\tilde e$}_{q_1} &=& 001~000~011~010~000 \\
\mbox{\boldmath $\tilde e$}_{q_2} &=& 000~101~101~010~000 \\
\mbox{\boldmath $\tilde e$}_{q_3} &=& 000~100~000~100~000 \\
\mbox{\boldmath $\tilde e$}_{q_4} &=& 001~001~110~100~000 .
\end{eqnarray}
Compare these error-paths with those in Fig.1. We observe that each error-path in Fig.2 (restricted to the section $[0, 4]$) is represented in Fig.1 in time-reversed order. That is, the original error-paths can be represented using the error-trellis associated with the corresponding reciprocal dual encoder.
\par
On the other hand, dividing the third column of $\tilde H_2(D)$ by $D$, we have the reduced canonical parity-check matrix
\begin{equation}
\tilde H_2'(D)\stackrel{\triangle}{=}\left(
\begin{array}{ccc}
1& 0 & 1 \\
D& 1+D & 0 
\end{array}
\right) .
\end{equation}
In this case [1], [8], error-paths associated with $\tilde H_2^T(D)$ can be represented using the error-trellis constructed based on $\tilde H_2^{'T}(D)$. Note that a factor $D^l$ in the column of $\tilde H(D)$ corresponds to backward-shifting by $l$ time units (i.e., $D^{-l}$) in terms of the original $H(D)$. This observation implies that error-trellis state-space reduction can be equally accomplished using backward-shifted error subsequences.

\section{QC Codes and Corresponding LDPC Convolutional Codes}
\subsection{LDPC Convolutional Codes Based on Circulant Matrices}
Each circulant in the parity-check matrix of a QC block code can be specified by a unique polynomial; the polynomial represents the entries in the first column of the circulant matrix. For example, a circulant matrix whose first column is $[1~1~1~0~1~0]^T$ is represented by the polynomial $1+D+D^2+D^4$. Using this correspondence, an LDPC convolutional code is constructed based on a parity-check matrix $H$ of a given QC code [10]. For example [10], let
\begin{equation}
H=\left(
\begin{array}{ccccc}
I_1& I_2& I_4& I_8& I_{16} \\
I_5& I_{10}& I_{20}& I_9& I_{18} \\
I_{25}& I_{19}& I_7& I_{14}& I_{28}
\end{array}
\right)
\end{equation}
be the parity-check matrix of a $(155, 64)$ QC code ($m=31$), where $I_x$ is a $31\times 31$ identity matrix with rows shifted cyclically to the left by $x$ positions. From $H$ we obtain the following parity-check matrix with polynomial entries:
\begin{equation}
H(D)=\left(
\begin{array}{ccccc}
D& D^2& D^4& D^8& D^{16} \\
D^5& D^{10}& D^{20}& D^9& D^{18} \\
D^{25}& D^{19}& D^7& D^{14}& D^{28}
\end{array}
\right) .
\end{equation}
({\it Remark:} It is stated [10] that the LDPC convolutional code is obtained by {\it unwrapping the constraint graph (i.e., Tanner graph)} of the QC code.) Note that the polynomials in $H(D)$ are all monomials. In this paper, we discuss exclusively using this specific example. However, the argument is entirely general.

\subsection{Reordering Rows of $H$ and the Corresponding $H(D)$}
Again, consider the parity-check matrix $H$ of the $(155, 64)$ QC code. Let us cyclically shift the first block of $m=31$ rows above by one position, the middle block of $31$ rows by five positions, and the last block of $31$ rows by $25$ positions. The resulting matrix is given by
\begin{equation}
H'=\left(
\begin{array}{ccccc}
I_0& I_1& I_3& I_7& I_{15} \\
I_0& I_5& I_{15}& I_4& I_{13} \\
I_0& I_{25}& I_{13}& I_{20}& I_3
\end{array}
\right) .
\end{equation}
Clearly, the QC block code and its associated constraint graph are unaffected by these row shifts. However, the convolutional code obtained based on the above procedure has the parity-check matrix
\begin{equation}
H'(D)=\left(
\begin{array}{ccccc}
1& D& D^3& D^7& D^{15} \\
1& D^5& D^{15}& D^4& D^{13} \\
1& D^{25}& D^{13}& D^{20}& D^3
\end{array}
\right) .
\end{equation}
We see that $H'(D)$ is not equivalent to $H(D)$. Two convolutional codes specified by $H(D)$ and $H'(D)$ are in fact different. We also remark that $H(D)$ and $H'(D)$ have different monomial entries and accordingly, when row/column factors are factored out, the resulting matrices have different OCLs.
\par
On the other hand, we have the following important fact [10]:
\par
{\it Property:} The LDPC convolutional codes obtained by unwrapping the constraint graph of the QC codes have their free distance $d_{free}$ lower-bounded by the minimum distance $d_{min}$ of the corresponding QC code.
\par
It is shown that the QC code associated with $H$ has a minimum distance $d_{min}=20$. Then $d_{free}$ of the convolutional code $C$ specified by $H(D)$ is lower-bounded by $d_{min}=20$. (It is conjectured that $C$ has a free distance $d_{free}$ of $24$ [10].) From the above property, we also have $d'_{free}\geq d_{min}=20$, where $d'_{free}$ is the free distance of the convolutional code $C'$ specified by $H'(D)$. In general, let $H'(D)$ be the parity-check matrix associated with $H'$, where $H'$ is the parity-check matrix obtained by applying cyclic shifts to the rows of each block of the original $H$. Above observations imply that the OCL of $H'(D)$ can be controlled to some extent with its free distance $d'_{free}$ being lower-bounded by the minimum distance $d_{min}$ of the QC code specified by $H$.

\section{Reduction of Overall Constraint Length}
\subsection{Row/Column Operations and Their Equivalent Representation}
Again, take the parity-check matrix $H$ given by Eq.(23). Here, let us cyclically shift the first block of $31$ rows above by one position. Then the first block $[I_1~I_2~I_4~I_8~I_{16}]$ changes to $[I_0~I_1~I_3~I_7~I_{15}]$. That is, the subscript number of each entry decreases by $1$. According to this change in $H$, the first row of $H(D)$ changes from $[D~D^2~D^4~D^8~D^{16}]$ to $[1~D~D^3~D^7~D^{15}]$. In terms of the power of $D$, this change is expressed as $[1~2~4~8~16]\rightarrow [0~1~3~7~15]$. In general, we observe that each entry decreases by one (modulo $31$) when we cyclically shift the rows above by one position. Continuing this procedure ({\it Remark:} it is assumed that row operations have been done), we see that the first row of $H(D)$ corresponds to one of the following five patterns in terms of the power of $D$:
\begin{equation}
\left\{
\begin{array}{l}
P_1=[0~1~3~7~15] \\
P_2=[30~0~2~6~14] \\
P_3=[28~29~0~4~12] \\
P_4=[24~25~27~0~8] \\
P_5=[16~17~19~23~0] . 
\end{array} \right.
\end{equation}
Similarly, the second and third rows of $H(D)$ are represented by
\begin{equation}
\left\{
\begin{array}{l}
Q_1=[0~5~15~4~13] \\
Q_2=[26~0~10~30~8] \\
Q_3=[16~21~0~20~29] \\
Q_4=[27~1~11~0~9] \\
Q_5=[18~23~2~22~0] 
\end{array} \right.\mbox{and}\left\{
\begin{array}{l}
R_1=[0~25~13~20~3] \\
R_2=[6~0~19~26~9] \\
R_3=[18~12~0~7~21] \\
R_4=[11~5~24~0~14] \\
R_5=[28~22~10~17~0] ,
\end{array} \right.
\end{equation}
respectively. Hence, when we apply cyclic shifts to the rows of each block of the original $H$, the resulting $H(D)$ can be specified by a pattern $[P_i, Q_j, R_k]^T$. For example, consider the pattern $[P_1, Q_1, R_3]^T$. The corresponding $H(D)$ (cf. Eq.(24)) is given by
\begin{equation}
H(D)=\left(
\begin{array}{ccccc}
1& D& D^3& D^7& D^{15} \\
1& D^5& D^{15}& D^4& D^{13} \\
D^{18}& D^{12}& 1& D^7& D^{21}
\end{array}
\right) .
\end{equation}
Since row operations have been done, let us apply column operations. Then we have
\begin{equation}
H'(D)=\left(
\begin{array}{ccccc}
1& 1& D^3& D^3& D^2 \\
1& D^4& D^{15}& 1& 1 \\
D^{18}& D^{11}& 1& D^3& D^8
\end{array}
\right) .
\end{equation}
We see that this is equivalent to the transformation from
\begin{equation}
S=\left(
\begin{array}{ccccc}
0& 1& 3& 7& 15 \\
0& 5& 15& 4& 13 \\
18& 12& 0& 7& 21
\end{array}
\right)
\end{equation}
to
\begin{equation}
S'=\left(
\begin{array}{ccccc}
0& 0& 3& 3& 2 \\
0& 4& 15& 0& 0 \\
18& 11& 0& 3& 8
\end{array}
\right) ,
\end{equation}
where subtraction is performed in each column of $S$ in order that the minimum is equal to zero. From $S'$, the OCL of the reduced $H'(D)$ is obtained as $3+15+18=36$.

\subsection{Reduction of Overall Constraint Length: Search Results}
\begin{figure}[tb]
\begin{center}
\includegraphics[width=7.5cm,clip]{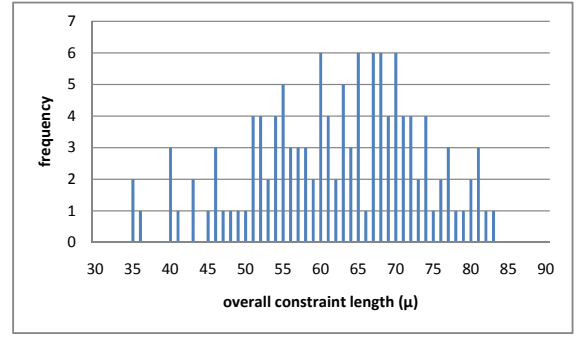}
\end{center}
\caption{Overall constraint length of a reduced $H'(D)$.}
\label{Fig.3}
\end{figure}
\begin{figure}[tb]
\begin{center}
\includegraphics[width=7.5cm,clip]{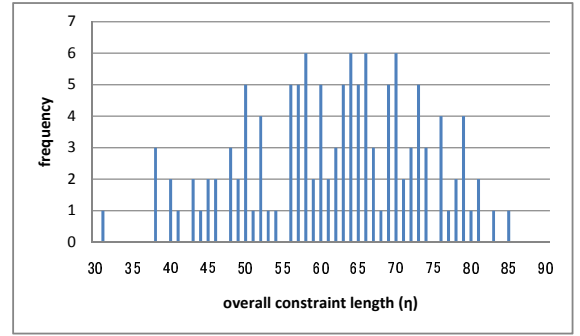}
\end{center}
\caption{Overall constraint length of a reduced $\tilde H'(D)$.}
\label{Fig.4}
\end{figure}

As we have seen in the previous section, there are $5 \times 5 \times 5=125$ patterns in total. By applying column operations to each pattern, we examined the OCL $\mu$ of the corresponding reduced parity-check matrix $H'(D)$. The result is shown in Fig.3, where the horizontal axis represents the OCL $\mu$ and the vertical axis represents its frequency. Observe that the minimum OCL $\mu_{min}$ is $35$, whereas the maximum OCL $\mu_{max}$ is $83$. That is, the values of $\mu$ cover a wide range. Next, based on the argument in Section II-B, we examined the OCL $\eta$ using the reciprocal dual encoder $\tilde H(D)$ associated with $H(D)$. The result is shown in Fig.4. We have $\eta_{min}=31$ and $\eta_{max}=85$. Note that in this example, the minimum OCL is further reduced using a reciprocal dual encoder. Moreover, after having applied row/column operations to each $H(D)$ (denote by $H'(D)$ the resulting matrix), we took its reciprocal version $\widetilde {H'}(D)$. Then again applying row/column operations to $\widetilde {H'}(D)$, we examined the OCL $\mu'$ of the resulting matrix. Using this method, we have further reduction with respect to the value of $\mu$. The result is shown in Fig.5. We observe that the maximum reduction $\Delta \mu(=\mu-\mu'\geq 0)$ of $16$ is realized. We remark that $\mu_{min}=35$ is reduced to $\mu'=34$ using this method.
\begin{figure}[tb]
\begin{center}
\includegraphics[width=7.5cm,clip]{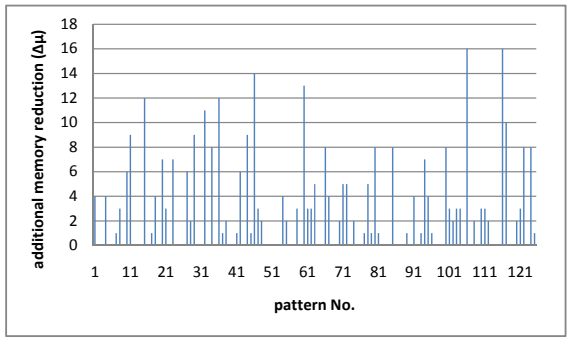}
\end{center}
\caption{Additional overall-constraint-length reduction.}
\label{Fig.5}
\end{figure}

\subsection{Efficient Search Method}
Our aim is the reduction of the OCL of a parity-check matrix $H(D)$. Since row operations have been done in a pattern $[P_i, Q_j, R_k]^T$, it is desirable for figures in each column to be close together (i.e., the difference $\delta_j$ between the maximum and the minimum in the $j$th column is small). In this case, each figure in the column becomes small after the column operation, which finally leads to the reduction of the OCL. Hence, we search for a pattern in which every column has $\delta_j \leq \Delta$, where $\Delta$ is a predetermined search parameter.
\par
For example, set $\Delta=20$. Consider the pattern $[P_2, Q_4, R_3]^T$:
\begin{equation}
S_1=\left(
\begin{array}{ccccc}
30& 0& 2& 6& 14 \\
27& 1& 11& 0& 9 \\
18& 12& 0& 7& 21
\end{array}
\right) .
\end{equation}
$\delta_j$ are given by $12$, $12$, $11$, $7$, and $12$, respectively and remain within $\Delta=20$. Applying column operations, we have
\begin{equation}
S_1'=\left(
\begin{array}{ccccc}
12& 0& 2& 6& 5 \\
9& 1& 11& 0& 0 \\
0& 12& 0& 7& 12
\end{array}
\right) .
\end{equation}
The OCL is given by $12+11+12=35$.
\par
Similarly, consider the pattern $[P_5, Q_5, R_5]^T$:
\begin{equation}
S_2=\left(
\begin{array}{ccccc}
16& 17& 19& 23& 0 \\
18& 23& 2& 22& 0 \\
28& 22& 10& 17& 0
\end{array}
\right) .
\end{equation}
In this case, $\delta_j$ are given by $12$, $6$, $17$, $6$, and $0$, respectively. Applying column operations, we have
\begin{equation}
S_2'=\left(
\begin{array}{ccccc}
0& 0& 17& 6& 0 \\
2& 6& 0& 5& 0 \\
12& 5& 8& 0& 0
\end{array}
\right) .
\end{equation}
Again, the OCL is given by $17+6+12=35$. Observe that these patterns have $\mu_{min}=35$.

\section{Conclusion}
In this paper, we examined the state-space complexity of the error-trellis of an LDPC convolutional code derived from the QC block code specified by a parity-check matrix $H$. Since the entries in the corresponding parity-check matrix $H(D)$ are all monomials, we can construct a reduced error-trellis using the method of [1] or that of [8]. We noticed that when cyclic shifts are applied to the rows of $H$, the QC code remains unchanged, whereas the corresponding parity-check matrix $H'(D)$, which defines another convolutional code, has row/column factors different from those in the original $H(D)$. That is, the OCL of $H''(D)$, where $H''(D)$ is the matrix obtained by factoring out row/column factors in $H'(D)$, varies depending on the row shifts applied to $H$. On the other hand, the free distance of the resulting convolutional code is still lower-bounded by the minimum distance of the original QC code. These facts imply that the state-space complexity of the error-trellis associated with $H'(D)$ can be controlled to some extent with the error-correction capability being preserved and this is our basic idea. By applying our method to the example in [10], we have shown that the OCL of the parity-check matrix of an LDPC convolutional code can be significantly reduced compared to the average one. The LDPC convolutional codes proposed by Tanner et al. have large constraint lengths. Therefore, it is stated [10] that the use of trellis-based decoding is not feasible. We basically agree on this point. However, it has been shown that an error-trellis with much lower state-space complexity than we imagined can be constructed, which gives some prospect of trellis-based decoding.

\section*{Acknowledgment}
This work was supported in part by the Japan Society for the Promotion of Science, under Grant-in-Aid No. 19500011.




%

\end{document}